\documentclass[a4paper]{article}

\usepackage{makecell}
\usepackage[utf8]{inputenc}
\usepackage[top=10mm]{geometry}
\usepackage{booktabs}
\usepackage[flushleft]{threeparttable}
\usepackage{authblk}
\usepackage{bm}
\usepackage{amsmath}
\usepackage{amssymb}
\usepackage{graphicx}
\usepackage[square,sort,numbers,compress]{natbib}
\usepackage{slashed}
\usepackage{float}
\usepackage{longtable,booktabs}
\usepackage{url}
\usepackage{relsize}
\usepackage[colorlinks,
linkcolor=blue,
anchorcolor=blue,
urlcolor=blue,
citecolor=blue
]{hyperref}
 \allowdisplaybreaks[4]

\title{\large \textbf{Meson-baryon scattering lengths without annihilation diagrams  to order $p^4$ in heavy baryon chiral perturbation theory}}

\author[1]{\small Jing Ou-Yang }
\author[1]{\small Ke Wang}
\author[1,2]{\small Bo-Lin Huang \thanks{blhuang@imu.edu.cn }}

\affil[1]{\textit{\smaller[0.50] School of Physical Science and Technology, $\&$ Inner Mongolia Key Laboratory of Microscale Physics and Atomic Manufacturing, Inner Mongolia University, Hohhot 010021, China}}
\affil[2]{\textit{\smaller[0.50] Research Center for Quantum Physics and Technologies, Inner Mongolia University, Hohhot 010021, China}}

\date{\small \today}
\begin{document}
\maketitle

%abstract
\begin{abstract}
We calculate the threshold $T$ matrices of the meson and baryon processes that have no annihilation diagrams: $\pi^{+}\Sigma^{+}$, $\pi^{+}\Xi^0$, $K^+p$, $K^+n$, and $\bar{K}^0\Xi^0$ to the fourth order in heavy baryon chiral perturbation theory. By performing least squares and Bayesian fits to the non-physical lattice QCD data, we determine the low-energy constants through both perturbative and non-perturbative iterative methods. By using these low-energy constants, we obtain the physical scattering lengths in these fits. The convergence behavior is not good across all channels in the perturbative method. The scattering lengths for the five channels, obtained by taking the median values from four different fitting approaches, are $a_{\pi^+\Sigma^+}=-0.16\pm 0.07\,\text{fm}$, $a_{\pi^+\Xi^0}=-0.04\pm0.04\,\text{fm}$, $a_{K^+p}=-0.41\pm 0.11\,\text{fm}$, $a_{K^+n}=-0.19\pm 0.10\,\text{fm}$, and $a_{\bar{K}^0\Xi^0}=-0.30\pm 0.07\,\text{fm}$, where the uncertainties are conservatively estimated by taking the maximum deviation between the median and extreme values of the statistical errors.

\end{abstract}

\section{Introduction}
The study of low-energy meson-baryon interactions represents a fundamental testing ground for quantum chromodynamics (QCD) in its non-perturbative regime. Among these processes, the precise determination of scattering lengths provides crucial insights into the dynamics of strong interactions at low energies, where the chiral symmetry of QCD plays an important role. Chiral perturbation theory (ChPT) \cite{Weinberg:1978kz,Gasser:1983yg,Leutwyler:1993iq,Scherer:2012xha} has emerged as the most systematic framework for analyzing such processes, offering a controlled expansion in powers of meson masses and momenta over the chiral symmetry breaking scale $\Lambda_\chi\simeq 1\,\text{GeV}$.

The development of ChPT for meson-baryon systems has progressed along two main approaches: the relativistic formulation \cite{Becher:1999he,Gegelia:1999gf,Fuchs:2003qc} and the nonrelativistic methods, specifically heavy baryon chiral perturbation theory (HBChPT) \cite{Jenkins:1990jv,Bernard:1992qa}. Relativistic approaches have achieved substantial progress in various aspects of meson-baryon interactions \cite{Schindler:2006it,Alarcon:2011zs,Chen:2012nx,Yao:2016vbz,Lu:2018zof}. Nevertheless, HBChPT remains particularly valuable for studying low-energy processes, where its nonrelativistic framework offers distinct advantages. This method allows us to systematically calculate scattering amplitudes while incorporating relativistic effects through $1/M_0$ corrections, where $M_0$ is the baryon mass in the chiral limit. Furthermore, at sufficiently low energies, where relativistic effects become negligible, HBChPT proves especially efficient and reliable for practical calculations.

Over the years, SU(2) HBChPT has been widely used to investigate the low-energy processes of pions and nucleons and has achieved many successes \cite{Fettes:1998ud,Fettes:2000xg,Krebs:2012yv,Entem:2014msa,Epelbaum:2014sza}, but its extension to SU(3) HBChPT introduces new complexities. The relatively large mass of strange quarks leads to slower convergence in the chiral expansion, making higher-order calculations particularly important. The third-order ($p^3$) calculations in SU(3) case can provide reasonably descriptions for some meson-baryon scattering processes from refs.~\cite{Kaiser:2001hr, Liu:2006xja, Liu:2007ct, Huang:2015ghe,Huang:2017bmx,Huang:2019not,Ou-Yang:2024neq}. A careful examination of the five specific channels without annihilation  was conducted in ref.~\cite{Torok:2009dg} through a combined analysis with lattice QCD data. It was emphasized that the SU(3) $\mathcal{O}(p^3)$ analysis suffers from unnaturally large (LECs) and poor convergence, making the expansion unreliable and preventing meaningful predictions for kaon-baryon scattering lengths. The study revealed that the third-order SU(3) expansion produced physically unreasonable results for these channels. The ``unreasonable" refers to the systematic instability and unnaturalness of the LECs, rather than to a specific scattering length. However, in our previous work \cite{Huang:2021fdt}, we performed the calculation of light meson-heavy meson scattering lengths to the fourth order within the SU(3) framework and obtained physically reasonable results through an iterative approach. Due to the large mass of the kaon, calculating up to the third order may lead to issues such as excessively large LECs and a complete breakdown of convergence. However, as in the calculation of light meson-heavy meson scattering lengths, we will extend the kaon-baryon scattering length calculation to the fourth order and employ iterative formulas. Within these frameworks, the problems caused by the large kaon mass will be eased, thereby allowing a reasonable result for the kaon-baryon scattering length predictions. Therefore, we will include the kaon-baryon scattering channels in our calculation.

Lattice QCD has achieved precision in meson-meson interaction studies \cite{Beane:2006gj,Beane:2007xs,Beane:2007uh}, with scattering length predictions accurate to within a few percent, owing to high signal-to-noise ratios and strong chiral symmetry constraints. However, baryonic systems present greater challenges, while pion-nucleon scattering with annihilation diagrams has been investigated \cite{Bulava:2022vpq}, the limited lattice data currently available for such processes remain statistically less reliable. In contrast, meson-baryon channels without annihilation diagrams (such as $\pi^+\Sigma^+$, $\pi^+\Xi^0$, $K^+p$, $K^+n$, and $\bar{K}^0\Xi^0$) continue to provide more reliable and computationally tractable results in lattice simulations \cite{Torok:2009dg}, as they avoid the complications of annihilation diagrams and maintain better signal quality.

In this work, we will calculate the threshold $T$ matrices of the meson and baryon processes that have no annihilation diagrams: $\pi^+\Sigma^+$, $\pi^+\Xi^0$, $K^+p$, $K^+n$, and $\bar{K}^0\Xi^0$ to the fourth order in HBChPT. By performing least squares and Bayesian fits to the non-physical lattice QCD data, we will determine the LECs through both perturbative and non-perturbative iterative methods. The Bayesian fitting approach offers the advantage of incorporating prior distributions for the LECs, where we can use natural size estimates and other prior information about these parameters. Bayesian methods have demonstrated some successful applications in chiral effective field theory studies \cite{Melendez:2017phj,Svensson:2021lzs,Svensson:2022kkj,Thim:2023fnl}. Considering the potential limitations in the available lattice QCD data, we will employ not only conventional least-squares fitting but also Bayesian fitting method, aiming to obtain more reliable and stable fitting results. Due to the relatively large mass of kaons, which may lead to convergence issues in the perturbative expansion, we additionally use a non-perturbative iterative approach for our fitting procedure as in our previous paper \cite{Huang:2021fdt}. Then, the physical scattering lengths will be obtained by extrapolating the corresponding parameters to their physical values. Through comprehensive extrapolation methods, we aim to obtain reliable and physically reasonable scattering length values at the physical point for these five channels.

The paper is structured as follows. Section~\ref{lagrangian} presents the theoretical framework and introduces the relevant chiral Lagrangians. In Section~\ref{tmatrices}, we derive and present the complete analytical expressions for the threshold $T$-matrices up to the fourth order in the chiral expansion. Section~\ref{scattering lengths} details our computational method for determining the scattering lengths from the obtained $T$-matrices. The main numerical results and their physical significance are analyzed and discussed in Section~\ref{results}. We conclude with a brief summary in Section~\ref{summary}. For completeness, the Appendix provides all necessary parameters for the $T$-matrix contributions arising from the fourth-order loop diagrams.

\section{Chiral Lagrangian}
\label{lagrangian}
In order to calculate the meson-baryon scattering lengths to the fourth order in heavy baryon chiral perturbation theory, the corresponding effective Lagrangian has the form
\begin{align}
\label{l1}
\mathcal{L}_{\text{eff}}=\mathcal{L}_{\phi\phi}+\mathcal{L}_{\phi B}.
\end{align}
The traceless Hermitian $3\times 3$ matrices $\phi$ and $B$ include the pseudoscalar Goldstone boson fields ($\pi$, $K$, $\bar{K}$, $\eta$) and
the octet-baryon fields ($N$, $\Lambda$, $\Sigma$, $\Xi$), respectively. The lowest-order SU(3) chiral Lagrangians for meson-meson and meson-baryon interactions take the form \cite{Gasser:1987rb,Krause:1990xc}
\begin{align}
\label{l2}
\mathcal{L}^{(2)}_{\phi\phi}=f^2\text{tr}(u_\mu u^\mu +\frac{\chi_{+}}{4}),
\end{align}
\begin{align}
\label{l3}
 \mathcal{L}_{\phi B}^{(1)}=\text{tr}(i\overline{B}[v\cdot D,B])+2D\,\text{tr}(\overline{B}S_{\mu}\{u^{\mu},B\})+2F\,\text{tr}(\overline{B}S_{\mu}[u^{\mu},B]),
\end{align}
where $f$ is the pseudoscalar decay constant in the chiral limit. The axial vector quantity $u^\mu=i\{\xi^{\dagger},\partial^\mu\xi\}/2$ contains odd number meson fields. The quantity $\chi_{\pm}=\xi^{\dagger}\chi\xi^{\dagger}\pm\xi\chi\xi$ with $\chi=\text{diag}(m_\pi^2,m_\pi^2,2m_K^2-m_\pi^2)$ introduces explicit chiral symmetry breaking terms. We choose the SU(3) matrix
\begin{align}
\label{l4}
U=\xi^2=\text{exp}(i\phi/f),
\end{align}
which collects the pseudoscalar Goldstone boson fields. Note that the so-called sigma parameterization was chosen in SU(2) HB$\chi$PT \cite{Mojzis:1997tu,Fettes:1998ud}. The $D_{\mu}$ denotes the chiral covariant derivative
\begin{align}
\label{l5}
[D_{\mu},B]=\partial_{\mu}B+[\Gamma_{\mu},B],
\end{align}
and $S_{\mu}$ is the covariant spin operator
\begin{align}
\label{l6}
S_\mu=\frac{i}{2}\gamma_5 \sigma_{\mu\nu}v^\nu,\quad S\cdot v=0,
\end{align}
\begin{align}
\label{l7}
\{S_\mu,S_\nu\}=\frac{1}{2}(v_\mu v_\nu-g_{\mu\nu}),\quad [S_\mu,S_\nu]=i\epsilon_{\mu\nu\sigma\rho}v^\sigma S^\rho,
\end{align}
where $\epsilon_{\mu\nu\sigma\rho}$ is the completely antisymmetric tensor in four indices, $\epsilon_{0123}=1.$ The chiral connection $\Gamma^\mu=[\xi^{\dagger},\partial^\mu\xi]/2$ contains even number meson fields. The physical values of the axial vector coupling constants $D$ and $F$ can be obtained from the mixed-action lattice QCD calculation of ref.~\cite{Lin:2007ap}. The complete second-order Lagrangian can be found in ref.~\cite{Muller:1996vy}. When only the threshold case is considered, it can be written as \cite{Liu:2006xja}
\begin{align}
\label{LphiB2}
\mathcal{L}_{\phi B}^{(2)}= &b_D\text{tr}(\overline{B}\{\chi_{+},B\})+b_F\text{tr}(\overline{B}[\chi_{+},B])+b_0\text{tr}(\overline{B}B)\text{tr}(\chi_{+})\nonumber\\&+\Big(2d_D+\frac{D^2-3F^2}{2M_0}\Big)\text{tr}(\overline{B}\{v\cdot u  v\cdot u,B\})+\Big(2d_F-\frac{DF}{M_0}\Big)\text{tr}(\overline{B}[v\cdot u  v\cdot u,B])\nonumber\\&+\Big(2d_0+\frac{F^2-D^2}{2M_0}\Big)\text{tr}(\overline{B}B)\text{tr}(v\cdot u  v\cdot u)+\Big(2d_1+\frac{3F^2-D^2}{3M_0}\Big)\text{tr}(\overline{B}v\cdot u)\text{tr}(v\cdot u B).
\end{align}
The first three terms proportional to the LECs $b_{D,F,0}$ give rise to explicit chiral symmetry breaking. The $1/M_0$ correction terms arise from the relativistic leading-order chiral Lagrangian, while the remaining terms are proportional to the LECs $d_D$, $d_F$, $d_0$, and $d_1$. The complete three-flavor Lorentz-invariant meson-baryon chiral Lagrangian at $\mathcal{O}(p^3)$, $\mathcal{L}_{\phi B}^{(3)}$, was systematically constructed in refs.~\cite{Oller:2006yh,Frink:2006hx,Oller:2007qd}. Only three independent terms contribute to the threshold behavior of meson-baryon scattering $T$-matrices, as given by \cite{Liu:2007ct}
\begin{align}
\label{LphiB3}
\mathcal{L}_{\phi B}^{(3)}=h_1 \text{tr}(\overline{B}B[\chi_{-},v\cdot u])+h_2\text{tr}(\overline{B}[\chi_{-},v\cdot u]B)+h_3\{\text{tr}(\overline{B}v\cdot u)\text{tr}(\chi_{-}B)-\text{tr}(\overline{B}\chi_{-})\text{tr}(v\cdot u B)\},
\end{align}
where $h_1$, $h_2$, and $h_3$  LECs that serve to absorb the divergences arising from loop diagrams.

We calculate the threshold $T$ matrices for the scattering of meson-baryons to the fourth order following the power counting rule \cite{Bernard:1995dp}. The dimension $n$ of any Feynman diagram is given by
\begin{align}
\label{l9}
    n=2L+1+\sum_{d}(d-2)N_{d}^{M}+\sum_{d}(d-1)N_{d}^{MB},
\end{align}
where $L$ is the number of loops, while $N_d^{M}$ and $N_d^{MB}$ denote the number of vertices with dimension $d$ from the pure mesonic, meson-baryon Lagrangians, respectively. The leading-order and next-to-leading-order contributions can be directly obtained from tree-level Lagrangians $\mathcal{L}_{\phi B}^{(1)}$ and $\mathcal{L}_{\phi B}^{(2)}$, respectively.

At both third and fourth orders, the amplitudes receive contributions from loop diagrams as well as tree-level terms involving $\mathcal{O}(p^3)$ and $\mathcal{O}(p^4)$ LECs derived from the meson-baryon chiral Lagrangians $\mathcal{L}_{\phi B}^{(3)}$ and $\mathcal{L}_{\phi B}^{(4)}$, respectively.  The full Lorentz-invariant $\mathcal{O}(p^4)$ Lagrangian $\mathcal{L}_{\phi B}^{(4)}$ has been established in ref.~\cite{Jiang:2016vax}. However, significant challenges remain: even when considering threshold $T$-matrices, many
$\mathcal{O}(p^4)$ LECs remain undetermined due to insufficient experimental constraints.

In this work, we focus exclusively on meson-baryon processes without annihilation diagrams. In the SU(3) flavor symmetry limit, we derive the following amplitude relations through isospin and U-spin symmetry \cite{Liu:2006xja}:
\begin{align}
\label{fivechannels}
T_{\pi^{+}\Sigma^{+}} = T_{K^{+}p} = T_{\bar{K}^0\Xi^0}, \quad
T_{\pi^{+}\Xi^0} = T_{K^{+}n}.
\end{align}
These symmetry relations allow us to parameterize all tree-level $\mathcal{O}(p^4)$ amplitudes using just two independent parameters. Consequently, a detailed analysis of $\mathcal{L}_{\phi B}^{(4)}$ is unnecessary for our present purposes.

\section{Threshold $T$-matrices}
\label{tmatrices}
\begin{figure}[t]
\centering
\includegraphics[height=9.2cm,width=8cm]{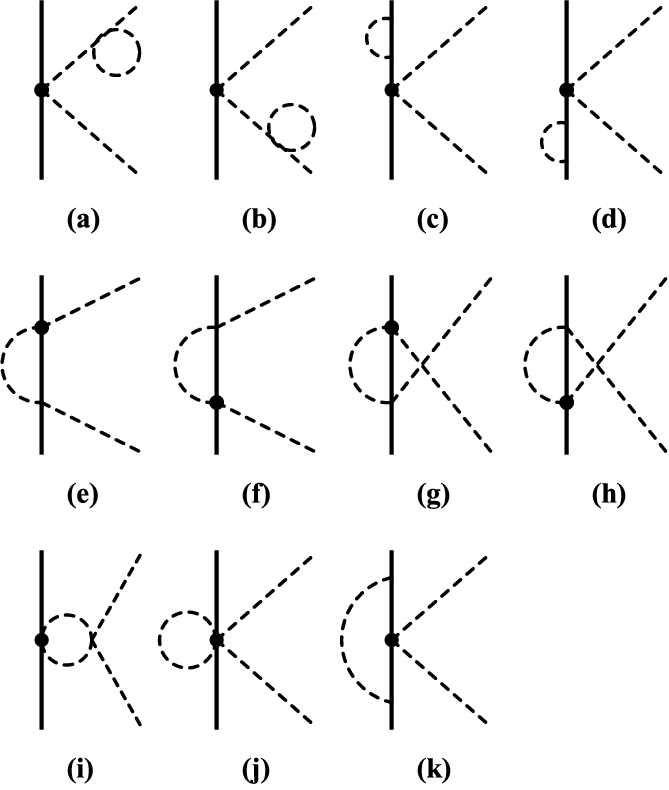}% Here is how to import EPS art
\caption{\label{fig:oneloopfeynman}
Nonvanishing one-loop diagrams contributing at chiral
order four. The heavy dots refer to vertices from $\mathcal{L}_{\phi B}^{(2)}$. Dashed lines represent Goldstone bosons and solid lines represent octet baryons.}
\end{figure}

We present the explicit expressions for the threshold $T$-matrices of the five meson-baryon processes listed in Eq.~(\ref{fivechannels}) to $\mathcal{O}(p^4)$ in the chiral expansion. The results through $\mathcal{O}(p^3)$ have been reported in refs.~\cite{Liu:2006xja,Liu:2007ct}. These elastic meson-baryon scattering processes are analyzed through isospin decomposition:
\begin{align}
\label{t1}
    T_{\pi^{+}\Sigma^{+}}=T_{\pi\Sigma}^{(2)},\,T_{\pi^{+}\Xi^{0}}=T_{\pi\Xi}^{(3/2)},\, T_{K^{+}p}=T_{KN}^{(1)},\, T_{K^{+}n}=\frac{1}{2}(T_{KN}^{(1)}+T_{KN}^{(0)}),\, T_{\bar{K}^{0}\Xi^{0}}=T_{\bar{K}\Xi}^{(1)}.
\end{align}
The fourth-order threshold $T$-matrices are determined as follows: tree-level terms are derived using the symmetry relations in Eq.~(\ref{fivechannels}), while loop corrections are calculated via the diagrams shown in Fig.~\ref{fig:oneloopfeynman}. Consequently, the threshold $T$-matrices with the four orders read
\begin{align}
\label{eq:tpiSigmaSum}
T_{\pi^{+}\Sigma^{+}}=\Big\{-\frac{m_\pi}{f_\pi^2}\Big\}+\Big\{\frac{m_\pi^2}{f_{\pi}^2}C_1\Big\}+\Big\{\frac{m_\pi^3}{f_{\pi}^2}\bar{h}_{123}+\mathcal{T}_{\pi^{+}\Sigma^{+}}^{\text{(N2LO)}}\Big\}+\Big\{\frac{m_{\pi}^4}{f_{\pi}^2}\bar{e}_1+\mathcal{T}_{\pi^{+}\Sigma^{+}}^{\text{(N3LO)}}\Big\},
\end{align}
\begin{align}
\label{eq:tpiXiSum}
T_{\pi^{+}\Xi^{0}}=\Big\{-\frac{m_\pi}{2f_\pi^2}\Big\}+\Big\{\frac{m_\pi^2}{2f_{\pi}^2}C_{01}\Big\}+\Big\{\frac{m_\pi^3}{f_{\pi}^2}\bar{h}_{1}+\mathcal{T}_{\pi^{+}\Xi^{0}}^{\text{(N2LO)}}\Big\}+\Big\{\frac{m_{\pi}^4}{f_{\pi}^2}\bar{e}_2+\mathcal{T}_{\pi^{+}\Xi^{0}}^{\text{(N3LO)}}\Big\},
\end{align}
\begin{align}
\label{eq:tKpSum}
T_{K^{+}p}=\Big\{-\frac{m_K}{f_K^2}\Big\}+\Big\{\frac{m_K^2}{f_{K}^2}C_1\Big\}+\Big\{\frac{m_K^3}{f_{K}^2}\bar{h}_{123}+\mathcal{T}_{K^{+}p}^{\text{(N2LO)}}\Big\}+\Big\{\frac{m_{K}^4}{f_{K}^2}\bar{e}_1+\mathcal{T}_{K^{+}p}^{\text{(N3LO)}}\Big\},
\end{align}
\begin{align}
\label{eq:tKnSum}
T_{K^{+}n}=\Big\{-\frac{m_K}{2f_K^2}\Big\}+\Big\{\frac{m_K^2}{2f_{K}^2}C_{01}\Big\}+\Big\{\frac{m_K^3}{f_{K}^2}\bar{h}_{1}+\mathcal{T}_{K^{+}n}^{\text{(N2LO)}}\Big\}+\Big\{\frac{m_{K}^4}{f_{K}^2}\bar{e}_2+\mathcal{T}_{K^{+}n}^{\text{(N3LO)}}\Big\},
\end{align}
\begin{align}
\label{eq:tKXiSum}
T_{\bar{K}^{0}\Xi^{0}}=\Big\{-\frac{m_K}{f_K^2}\Big\}+\Big\{\frac{m_K^2}{f_{K}^2}C_1\Big\}+\Big\{\frac{m_K^3}{f_{K}^2}\bar{h}_{123}+\mathcal{T}_{\bar{K}^{0}\Xi^{0}}^{\text{(N2LO)}}\Big\}+\Big\{\frac{m_{K}^4}{f_{K}^2}\bar{e}_1+\mathcal{T}_{\bar{K}^{0}\Xi^{0}}^{\text{(N3LO)}}\Big\},
\end{align}
where we have defined $C_{01}=C_0+C_1$, and $C_{0,1}$ are given by
\begin{align}
\label{c0c1}
    C_0=4b_F-4b_0-b_2+b_3-\frac{b_4}{2},\quad C_1=-4b_D-4b_0+b_1+b_3+\frac{b_4}{2},
\end{align}
with
\begin{align}
\label{b1tob4}
    b_1=2d_D+\frac{D^2-3F^2}{2M_0},\,
    b_2=2d_F-\frac{DF}{M_0},\,
    b_3=2d_0+\frac{F^2-D^2}{2M_0},\,
    b_4=2d_1+\frac{3F^2-D^2}{3M_0}.
\end{align}
Although the LECs are presented in combination form in the second-order $T$-matrices, the LECs $b_{D,F,0,1,2,3,4}$ from the second-order Lagrangian will appear in a split form in the $T$-matrices derived from the fourth-order loop diagrams. Consequently, the final data fitting will be performed using this split configuration. Furthermore, we have $\bar{h}_{123}=4(h_1^{\text{r}}-h_2^{\text{r}}+h_3)$ and $\bar{h}_{1}=4h_1^{\text{r}}$, where the scale-independent LECs are used. Thus, all terms $\text{ln}(m_{\pi,K,\eta}/\mu)$ have disappeared in our final expressions. In addition, the scale-independent LECs $\bar{e}_{1,2}$ are also used in the fourth order $T$-matrices.

The loop-diagram contributions to the third-order $T$-matrices, $\mathcal{T}_{\phi B}^{(\text{N2LO})}$, are explicitly derived in ref.~\cite{Liu:2006xja}. After performing scale-independent renormalization, we have
\begin{align}
\label{tpisigman2lo}
    \mathcal{T}_{\pi^+\Sigma^+}^{(\text{N2LO})}=\frac{m_\pi^2}{8\pi^2 f_\pi^4}\Big\{-\frac{3}{2}m_\pi-\sqrt{m_K^2-m_\pi^2}\arccos\frac{m_\pi}{m_K}+\frac{\pi}{2}\Big[3F^2m_\pi-\frac{1}{3}D^2m_\eta\Big]\Big\},
\end{align}
\begin{align}
\label{tpixin2lo}
    \mathcal{T}_{\pi^+\Xi^0}^{(\text{N2LO})}=&\frac{m_\pi^2}{16\pi^2 f_\pi^4}\Big\{-\frac{3}{2}m_\pi-\sqrt{m_K^2-m_\pi^2}\Big(\pi+\arccos\frac{m_\pi}{m_K}\Big)\nonumber\\
    &+\frac{\pi}{4}\Big[3(D-F)^2m_\pi-\frac{1}{3}(D+3F)^2m_\eta\Big]\Big\},
\end{align}
\begin{align}
\label{tkpn2lo}
    \mathcal{T}_{K^+p}^{(\text{N2LO})}=&\frac{m_K^2}{16\pi^2 f_K^4}\Big\{-3 m_K+2\sqrt{m_K^2-m_\pi^2}\ln\frac{m_K+\sqrt{m_K^2-m_\pi^2}}{m_\pi}-3\sqrt{m_\eta^2-m_K^2}\arccos\frac{m_K}{m_\eta}\nonumber\\
    &-\frac{\pi}{6}(D-3F)\Big[2(D+F)\frac{m_\pi^2}{m_\eta+m_\pi}+(D+5F)m_\eta\Big]\Big\},
\end{align}
\begin{align}
\label{tknn2lo}
    \mathcal{T}_{K^+n}^{(\text{N2LO})}=&\frac{m_K^2}{16\pi^2 f_K^4}\Big\{-\frac{3}{2} m_K+\frac{5}{2}\sqrt{m_K^2-m_\pi^2}\ln\frac{m_K+\sqrt{m_K^2-m_\pi^2}}{m_\pi}-\frac{3}{2}\sqrt{m_\eta^2-m_K^2}\arccos\frac{m_K}{m_\eta}\nonumber\\
    &+\frac{\pi}{3}(D-3F)\Big[(D+F)\frac{m_\pi^2}{m_\eta+m_\pi}+\frac{1}{2}(3D-F)m_\eta\Big]\Big\},
\end{align}
\begin{align}
\label{tk0barxin2lo}
    \mathcal{T}_{\bar{K}^0\Xi^0}^{(\text{N2LO})}=&\frac{m_K^2}{16\pi^2 f_K^4}\Big\{-3 m_K+2\sqrt{m_K^2-m_\pi^2}\ln\frac{m_K+\sqrt{m_K^2-m_\pi^2}}{m_\pi}-3\sqrt{m_\eta^2-m_K^2}\arccos\frac{m_K}{m_\eta}\nonumber\\
    &-\frac{\pi}{6}(D+3F)\Big[2(D-F)\frac{m_\pi^2}{m_\eta+m_\pi}+(D-5F)m_\eta\Big]\Big\}.
\end{align}

At the fourth order $\mathcal{O}(p^4)$, one also has $T$-matrices from one-loop diagrams generated by the vertices of $\mathcal{L}_{\phi B}^{(2)}$, $\mathcal{L}_{\phi B}^{(1)}$ and $\mathcal{L}_{\phi \phi}^{(2)}$ are shown in Fig.~\ref{fig:oneloopfeynman}. After performing tedious calculations, the loop-diagram contributions to the fourth-order $T$-matrices, $\mathcal{T}_{\phi B}^{(\text{N3LO})}$, are given by
\begin{align}
\label{tpisigman3lo}
      \mathcal{T}_{\pi^+\Sigma^+}^{(\text{N3LO})}=\frac{1}{\pi^2 f_\pi^4}\Big\{\alpha_{1}m_\pi^4+\beta_{1}m_K^4+\gamma_{1}m_\pi^2m_K^2+\delta_{1}\sqrt{m_K^2-m_\pi^2}\arccos\frac{m_\pi}{m_K}\Big\},
\end{align}
\begin{align}
\label{tpixi0n3lo}
      \mathcal{T}_{\pi^+\Xi^0}^{(\text{N3LO})}=&\frac{1}{\pi^2 f_\pi^4}\Big\{\alpha_{2}m_\pi^4+\beta_{2}m_K^4+\gamma_{2}m_\pi^2m_K^2+\delta_{2}\sqrt{m_K^2-m_\pi^2}\arccos\frac{m_\pi}{m_K}\nonumber\\
      &+\epsilon_2\sqrt{m_K^2-m_\pi^2}\arccos\frac{-m_\pi}{m_K}\Big\},
\end{align}
\begin{align}
\label{tkpn3lo}
      \mathcal{T}_{K^+p}^{(\text{N3LO})}=&\frac{1}{\pi^2 f_K^4}\Big\{\alpha_{3}m_\pi^4+\beta_{3}m_K^4+\gamma_{3}m_\pi^2m_K^2+\delta_3\sqrt{m_\eta^2-m_K^2}\arccos\frac{m_K}{m_\eta}\nonumber\\
      &+\epsilon_3\sqrt{m_K^2-m_\pi^2}\ln\frac{m_K+\sqrt{m_K^2-m_\pi^2}}{m_\pi}\Big\},
\end{align}
\begin{align}
\label{tknn3lo}
      \mathcal{T}_{K^+n}^{(\text{N3LO})}=&\frac{1}{\pi^2 f_K^4}\Big\{\alpha_{4}m_\pi^4+\beta_{4}m_K^4+\gamma_{4}m_\pi^2m_K^2+\delta_4\sqrt{m_\eta^2-m_K^2}\arccos\frac{m_K}{m_\eta}\nonumber\\
      &+\epsilon_4\sqrt{m_K^2-m_\pi^2}\ln\frac{m_K+\sqrt{m_K^2-m_\pi^2}}{m_\pi}\Big\},
\end{align}
\begin{align}
\label{tkbar0xi0n3lo}
      \mathcal{T}_{\bar{K}^0 \Xi^0}^{(\text{N3LO})}=&\frac{1}{\pi^2 f_K^4}\Big\{\alpha_{5}m_\pi^4+\beta_{5}m_K^4+\gamma_{5}m_\pi^2m_K^2+\delta_5\sqrt{m_\eta^2-m_K^2}\arccos\frac{m_K}{m_\eta}\nonumber\\
      &+\epsilon_5\sqrt{m_K^2-m_\pi^2}\ln\frac{m_K+\sqrt{m_K^2-m_\pi^2}}{m_\pi}\Big\},
\end{align}
where the parameters $\alpha_i$, $\beta_i$, $\gamma_i$, $\delta_i$ and $\epsilon_i$ ($i=1, 2, 3, 4, 5$) can be found in Appendix~\ref{parameters}. The Gell-Mann-Okubo (GMO) relation $m_\eta=\sqrt{(4m_K^2-m_\pi^2)/3}$ has also been used to simplify the equations.

\section{Scattering lengths}
\label{scattering lengths}
The $S$-wave scattering length is defined by
\begin{align}
\label{eq:pscatteringlength}
    a=\frac{1}{4\pi}\frac{M_B}{m_\phi+M_B}T_{\phi B},
\end{align}
where $m_\phi$ is the meson mass and $M_B$ is the baryon mass. We take the sign convention where a repulsive interaction corresponds to a negative scattering length. The chiral expansion may not converge when truncated at finite order, introducing uncontrolled systematic errors. For reliable predictions, the $T$-matrix can be iterated to infinite order. As in the phenomenological resummation approach of ref.~\cite{Kaiser:1995eg}, we use a Lippmann-Schwinger equation with a momentum cutoff scale $\mu$ to obtain finite results. For a single-channel separable potential, which is the case considered here since the five channels do not couple to others within the octet, the scattering length is given by
\begin{align}
\label{eq:iscatteringlength}
    a=a_{\text{Born}}\Big(1-\frac{1}{2}\mu\, a_{\text{Born}}\Big)^{-1}.
\end{align}
Here, $a_{\text{Born}}$ incorporates contributions from all diagrams with the exception of the iterated diagrams (e) and (f) in Fig.~\ref{fig:oneloopfeynman} of this work, as well as the fourth diagram in Fig.~1 of ref.~\cite{Liu:2006xja}. We only need to subtract the contributions from these three iterated diagrams where the internal and external mesons coincide, which can be obtained through the iteration of first- and second-order tree diagrams. In our convention, these contributions can be derived by multiplying $m_\phi/(8\pi^2)$ with the corresponding first- and second-order tree-level amplitudes. Thus, our calculation includes all contributions from tree, renormalization, and crossed diagrams up to the fourth order in the potential iteration. As a result, the scattering lengths do not explicitly include the imaginary parts. However, resummation may give rise to resonance states, for which unitarity demands a corresponding imaginary part. Nevertheless, the generation of an imaginary part through this resummation method is neither expected nor physically relevant. It should be noted that the iterative form will obscure the power-counting scheme.

\section{Results and discussion}
\label{results}
Before making predictions, we need to determine the LECs. We employ two distinct approaches to constrain these parameters: a conventional least-squares fitting method and a Bayesian statistical analysis framework.

\subsection{Least-squares fitting}

We determine the LECs $b_{D,F,0,1,2,3,4}$, $\bar{h}_{1}$, $\bar{h}_{123}$, and $\bar{e}_{1,2}$ through a least-squares fit to lattice QCD data. The goodness-of-fit is quantified by the $\chi^2$ function \cite{Dobaczewski:2014jga,Carlsson:2015vda}:
\begin{align}
\label{chi2}
    \chi^2=\sum_{i=1}^{N_d}\frac{(a_{i}^{\text{ChPT}}-a_{i}^{\text{Lattice}})^2}{\Delta a_{i}^2},
\end{align}
where $a_i^{\text{ChPT}}$ denotes the scattering lengths calculated from our chiral expansions, $a_i^{\text{Lattice}}$ represents the corresponding lattice QCD results from ref.~\cite{Torok:2009dg}, and $\Delta a_i$ is the combined uncertainty calculated by adding in quadrature the statistical and systematic uncertainties from the lattice QCD results. We use lattice data for five scattering processes: $\pi^{+}\Sigma^{+}$, $\pi^{+}\Xi^{0}$, $K^{+}p$, $K^{+}n$, and $\bar{K}^{0}\Xi^{0}$, including their scattering lengths, hadron masses and decay constants, see Table~\ref{tab:data}. We also use the Gell-Mann-Okubo (GMO) mass relation and $f_\eta = 1.3 f_\pi$ in all calculations.

The lattice data are calculated using unphysical quark masses, which leads to a larger mass for the light meson ($\sim 700$ MeV). However, the chiral expansions are based on the ratio $m_\phi/\Lambda_\chi$, where $m_\phi$ is the light meson mass and $\Lambda_{\chi}=1/(4\pi f_\pi)$ is the chiral symmetry breaking scale. The large meson mass might cause convergence issues in these expansions. Additionally, the predicted scattering length should not depend on the model choices. To address this, we perform two analyses: Fit LSp implements a fully perturbative treatment using Eq.~(\ref{eq:pscatteringlength}), while Fit LSu employs a non-perturbative approach based on Eq.~(\ref{eq:iscatteringlength}). This allows us to check the consistency of our predictions.
\begin{table*}[!t]
\centering
\begin{threeparttable}
\caption{\label{tab:data}The lattice data of the hadron masses, decay constants, and scattering lengths from ref.~\cite{Torok:2009dg}.}
\renewcommand{\arraystretch}{1.3}
\begin{tabular}{cccccccc}
\toprule \toprule
 $m_\pi$ (MeV) & $290.20\pm 0.49$ & $352.10\pm 0.41$ & $489.84\pm 1.62$ & $592.17\pm 0.72$ & \\
 $m_K$ (MeV) & $580.64\pm 0.83$ & $596.95\pm 0.45$ & $639.48\pm 0.78$ & $680.22\pm 1.07$ & \\
 $m_N$ (MeV) & $1101.52\pm 9.71$ & $1156.14\pm 5.14$ & $1273.74\pm 4.11$ & $1379.82\pm 2.65$ & \\
 $m_\Sigma$ (MeV) & $1324.41\pm 3.51$ & $1346.67\pm 3.25$ & $1393.87\pm 3.91$ & $1454.33\pm 2.11$ & \\
 $m_\Xi$ (MeV) & $1400.50\pm 4.29$ & $1422.12\pm 3.39$ & $1457.48\pm 2.91$ & $1493.48\pm 2.41$ & \\
 $f_\pi$ (MeV) & $103.33\pm 0.18$ & $107.16\pm 0.16$ & $113.94\pm 0.16$ & $120.14\pm 0.36$ & \\
 $f_K$ (MeV) & $119.81\pm 0.11$ & $120.34\pm 0.20$ & $122.52\pm 0.19$ & $125.61\pm 0.35$ & \\
\midrule
 $a_{\pi\Sigma}$ (fm) & $-0.27\pm 0.02$  & $-0.30\pm 0.02$  &  $-0.29\pm 0.03$ & $-0.30\pm 0.03$ & \\
 $a_{\pi\Xi}$ (fm) & $-0.14\pm 0.02$  & $-0.15\pm 0.03$  &  $-0.14\pm 0.02$ & $-0.15\pm 0.03$ & \\
 $a_{Kp}$ (fm) & $-0.35\pm 0.07$  & $-0.37\pm 0.04$  &  $-0.29\pm 0.13$ & $-0.28\pm 0.06$ & \\
 $a_{Kn}$ (fm) & $-0.18\pm 0.06$  & $-0.17\pm 0.05$  &  $-0.13\pm 0.05$ & $-0.11\pm 0.05$ & \\
 $a_{K\Xi}$ (fm) & $-0.33\pm 0.03$  & $-0.35\pm 0.03$  &  $-0.27\pm 0.08$ & $-0.29\pm 0.06$ & \\
\bottomrule \bottomrule
\end{tabular}
\end{threeparttable}
\end{table*}

\begin{table*}[!t]
\centering
\resizebox{\textwidth}{!}{%
\begin{threeparttable}
\caption{\label{tab:consLSfitting}Values of the various fits with the
correlations between the parameters through the least-squares fitting method. For a detailed
description of these fits, see the main text.}
\renewcommand{\arraystretch}{1.3}
\begin{tabular}{ccccccccccccccccccccccccccccccccc}
\toprule \toprule
 & Fit LSp & $b_D$ & $b_F$ & $b_0$ & $b_1$ & $b_2$ & $b_3$ & $b_4$ &  $\bar{h}_{1}$ & $\bar{h}_{123}$ & $\bar{e}_1$ & $\bar{e}_2$ & \\
\midrule

$b_D\,(\text{GeV}^{-1})$ &$0.38\pm 0.25$&$1.00$&$\hspace{-0.7em}-0.50$ & $0.96$& $0.96$ & $\hspace{-0.7em}-0.34$ &$0.98$&$\hspace{-0.7em}-0.76$& $0.47$ & $0.47$ & $\hspace{-0.7em}-0.21$ & $\hspace{-0.7em}-0.61$ &\\

$b_F\,(\text{GeV}^{-1})$ &$\hspace{-0.7em}-0.44\pm 0.16$&  &$1.00$&$\hspace{-0.7em}-0.32$& $\hspace{-0.7em}-0.28$ &$0.97$ &$\hspace{-0.7em}-0.46$ &$\hspace{-0.7em}-0.12$ & $0.44$ & $0.48$& $\hspace{-0.7em}-0.56$ & $\hspace{-0.7em}-0.26$ &\\

$b_0\,(\text{GeV}^{-1})$ &$2.49\pm 0.87$&  &   &$1.00$&$0.98$&$\hspace{-0.7em}-0.13$&$0.98$&$\hspace{-0.7em}-0.88$ & $0.61$ & $0.58$ & $\hspace{-0.7em}-0.26$ & $\hspace{-0.7em}-0.72$ &\\

$b_1\,(\text{GeV}^{-1})$&$2.92\pm 1.66$&    &    & & $1.00$&$\hspace{-0.7em}-0.09$&$0.94$&$\hspace{-0.7em}-0.91$& $0.65$ & $0.65$ & $\hspace{-0.7em}-0.36$ & $\hspace{-0.7em}-0.75$ &\\

$b_2\,(\text{GeV}^{-1})$ &$\hspace{-0.7em}-1.94\pm 0.72$&    &    & & & $1.00$& $\hspace{-0.7em}-0.29$& $\hspace{-0.7em}-0.32$& $0.58$ & $0.63$ & $\hspace{-0.7em}-0.65$ & $\hspace{-0.7em}-0.42$ &\\

$b_3\,(\text{GeV}^{-1})$ &$8.42\pm 2.97$&    &    & & & & $1.00$& $\hspace{-0.7em}-0.78$& $0.48$ & $0.44$ &$\hspace{-0.7em}-0.12$ & $\hspace{-0.7em}-0.62$ &\\

$b_4\,(\text{GeV}^{-1})$&$\hspace{-0.7em}-1.13\pm 1.50$&    &  & & & & & $1.00$& $\hspace{-0.7em}-0.86$ & $\hspace{-0.7em}-0.87$& $0.58$ & $0.88$ &\\

$\bar{h}_1\,(\text{GeV}^{-2})$     &$3.98\pm 1.66$&    &    & & & & & & $1.00$& $0.92$ & $\hspace{-0.7em}-0.76$ & $\hspace{-0.7em}-0.98$ &\\

$\bar{h}_{123}\,(\text{GeV}^{-2})$ &$1.35\pm 2.14$&    &  & &   & & & & & $1.00$& $\hspace{-0.7em}-0.90$ & $\hspace{-0.7em}-0.87$ &\\

$\bar{e}_1\,(\text{GeV}^{-3})$     &$0.45\pm 1.51$&    &   & & &  & & & & & $1.00$& $0.68$ &\\

$\bar{e}_2\,(\text{GeV}^{-3})$     &$\hspace{-0.7em}-4.39\pm 2.28$&    & & & & &    & & & & & $1.00$&\\
$\chi^2/\text{d.o.f.}$&$\frac{3.21}{20-11}=0.4$&    &    &\\

\midrule
\midrule

& Fit LSu & $b_D$ & $b_F$ & $b_0$ & $b_1$ & $b_2$ & $b_3$ & $b_4$ &  $\bar{h}_{1}$ & $\bar{h}_{123}$ & $\bar{e}_1$ & $\bar{e}_2$ & $\mu$ & \\
\midrule

$b_D\,(\text{GeV}^{-1})$           &$1.51\pm 0.33$&$1.00$ &$0.32$ & $\hspace{-0.7em}-0.10$& $0.05$ & $\hspace{-0.7em}-0.63$ &$\hspace{-0.7em}-0.49$&$0.33$& $0.72$ & $\hspace{-0.7em}-0.17$ & $\hspace{-0.7em}-0.68$ & $0.86$ &  $0.11$ &\\

$b_F\,(\text{GeV}^{-1})$           &$\hspace{-0.7em}-0.72\pm 0.31$&  &$1.00$&$\hspace{-0.7em}-0.16$& $0.89$ &$0.46$ &$\hspace{-0.7em}-0.15$ &$0.92$ & $\hspace{-0.7em}-0.20$ & $0.40$ & $0.12$ & $0.15$ & $\hspace{-0.7em}-0.52$ &\\

$b_0\,(\text{GeV}^{-1})$           &$4.63\pm 1.12$&  &   &$1.00$&$0.09$&$0.23$&$0.90$&$\hspace{-0.7em}-0.23$ & $\hspace{-0.7em}-0.47$& $0.77$ & $0.61$ & $\hspace{-0.7em}-0.39$ &  $\hspace{-0.7em}-0.69$ &\\

$b_1\,(\text{GeV}^{-1})$           &$5.11\pm 3.19$&    &    & & $1.00$ & $0.70$&$0.19$&$0.90$& $\hspace{-0.7em}-0.53$ & $0.62$ & $0.40$ &  $\hspace{-0.7em}-0.05$&  $\hspace{-0.7em}-0.71$ &\\

$b_2\,(\text{GeV}^{-1})$           &$\hspace{-0.7em}-4.11\pm 1.69$&    &    & & & $1.00$& $0.57$& $0.38$& $\hspace{-0.7em}-0.95$ & $0.68$ & $0.86$ & $\hspace{-0.7em}-0.71$& $\hspace{-0.7em}-0.70$  &\\

$b_3\,(\text{GeV}^{-1})$           &$16.43\pm 3.71$&    &    & & & & $1.00$& $\hspace{-0.7em}-0.22$& $\hspace{-0.7em}-0.79$ & $0.82$& $0.87$ & $\hspace{-0.7em}-0.71$ &  $\hspace{-0.7em}-0.73$ &\\

$b_4\,(\text{GeV}^{-1})$           &$0.24\pm 4.15$&    &     & & & & & $1.00$& $\hspace{-0.7em}-0.14$ & $0.27$ &  $\hspace{-0.7em}-0.02$ &  $0.33$ & $\hspace{-0.7em}-0.40$  &\\

$\bar{h}_1(\text{GeV}^{-2})$     &$8.00\pm 3.98$&    &     & & & & & & $1.00$ & $\hspace{-0.7em}-0.75$ & $\hspace{-0.7em}-0.95$ & $0.82$ &  $0.72$  &\\

$\bar{h}_{123}\,(\text{GeV}^{-2})$ &$\hspace{-0.7em}-1.36\pm 8.43$&    &     & & & & & & & $1.00$ & $0.82$ & $\hspace{-0.7em}-0.51$&  $\hspace{-0.7em}-0.98$  &\\

$\bar{e}_1\,(\text{GeV}^{-3})$     &$8.75\pm 5.95$&    &     & & & & &  & & & $1.00$ & $\hspace{-0.7em}-0.90$ & $\hspace{-0.7em}-0.76$&\\

$\bar{e}_2\,(\text{GeV}^{-3})$     &$\hspace{-0.7em}-9.27\pm 4.08$&    &     & & & & & & & & & $1.00$& $0.43$ &\\

$\mu\,(\text{GeV})$     &$0.72\pm 0.11$&    &     & & & & & & & & & & $1.00$&\\

$\chi^2/\text{d.o.f.}$&$\frac{1.50}{20-12}=0.2$&    &    &\\

\bottomrule \bottomrule
\end{tabular}
\end{threeparttable}}%
\end{table*}

\begin{figure}[!t]
\centering
\includegraphics[height=14.30cm,width=13.5cm]{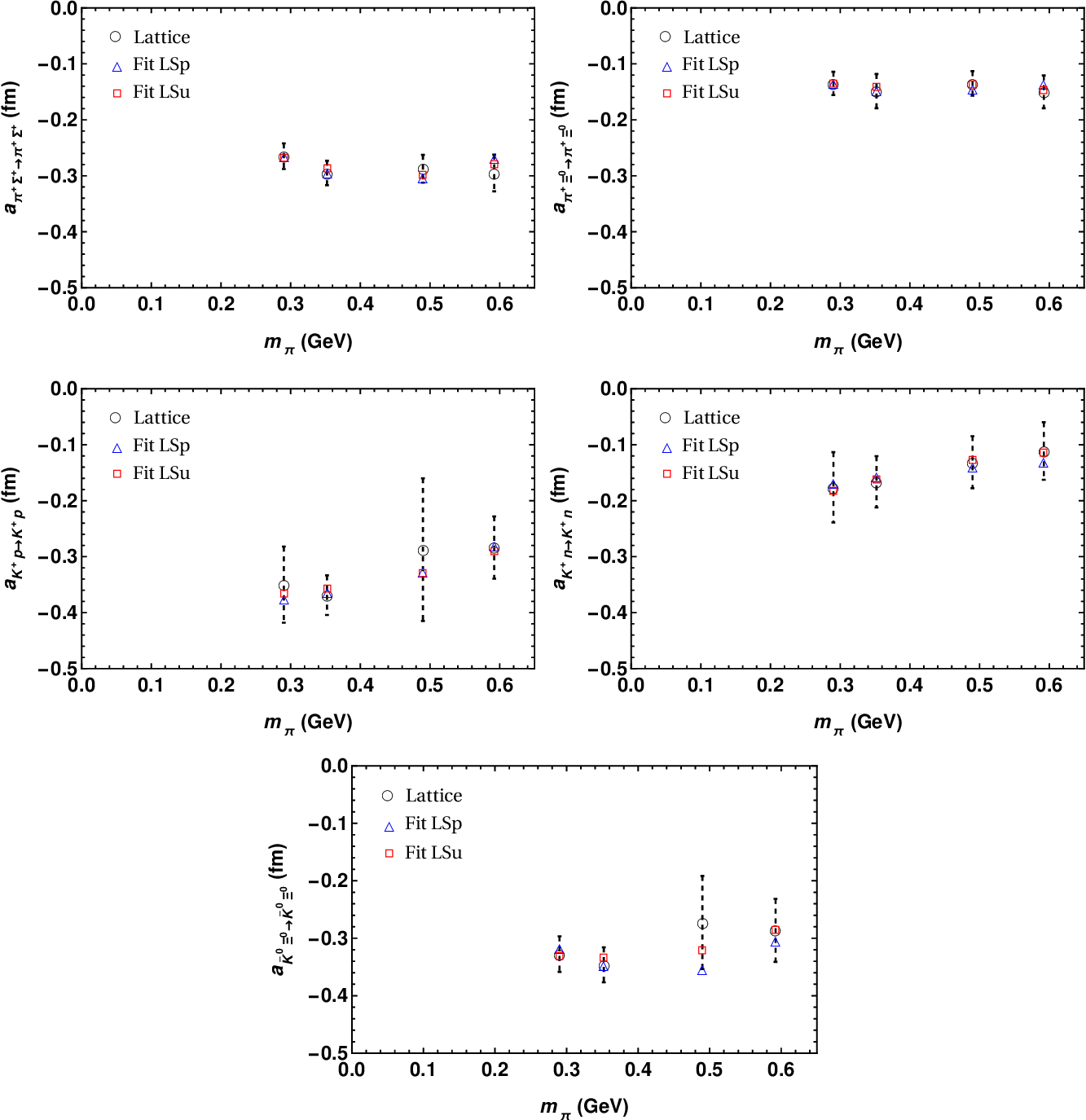}% Here is how to import EPS art
\caption{\label{fig:fitLS}Comparison between the five-channel fitting results obtained using the least-squares fitting method and the lattice QCD data for scattering lengths reported in ref.~\cite{Torok:2009dg}.}
\end{figure}

The resulting LECs with the correlations between the parameters for the two fits can be found in Table~\ref{tab:consLSfitting}. The LEC values from the LSp and LSu fits show some differences because the LSu fit uses iterations of the amplitudes, while the LSp fit does not. The LSu fit has more parameters and fits the data better, giving a chi-squared value per degree of freedom ($\chi^2/\text{d.o.f}$) that is half of the LSp fit's value. Both $\chi^2/\text{d.o.f.}$ values being below one suggests potential overfitting or an overestimation of lattice data errors. The errors in the lattice data include both statistical and systematic components from ref.~\cite{Torok:2009dg}, and may particularly overestimate the uncertainties in the scattering lengths of kaons and baryons. To address potential overfitting, we will employ a Bayesian fitting method with parameter priors in subsequent analyses. Examining the uncertainties of the LECs reveals that each parameter's statistical uncertainty reflects its permissible variation while maintaining a good fit. However, these LECs cannot vary independently due to their strong mutual correlations. Consequently, the large LEC uncertainties in our fits do not necessarily translate to significant errors in scattering lengths, since proper error analysis requires the full covariance matrix. The limited amount of fitted data relative to the number of parameters also contributes to these large uncertainties. In addition, the LSp fit exhibits smaller LEC uncertainties than the LSu fit, attributable to the additional parameter $\mu$ in the LSu fit and the non-linear behavior in certain regions that can introduce substantial errors if the parameters fall within them. Regarding the correlations among the LECs, we find that the iterative fit (LSu) reduces the parameter correlations compared to the perturbative fit (LSp). In the LSp fit, parameter pairs such as ($b_D$, $b_0$), ($b_D$, $b_1$) and ($b_D$, $b_3$) exhibit strong correlations, whereas the LSu fit weakens these dependencies. Despite the large uncertainties and strong correlations among the LECs, their values remain mostly within the natural size. However, the approximate agreement between LSp and LSu does not strictly resolve the convergence issues. Instead, it indicates that higher-order effects are numerically unimportant and that the results are relatively insensitive to the fitting scheme.

The corresponding scattering lengths from the two fits are shown in Fig.~\ref{fig:fitLS}. The scattering lengths obtained from both fitting approaches show overall good agreement with lattice QCD data in all five channels. For the $K^{+}p$, the value of the scattering length from lattice QCD has large error ($\sim 44\%$) at $m_\pi\simeq 490$ MeV. However, the scattering lengths obtained from both fits remain equally close to the central values of the lattice QCD, indicating that these large uncertainties do not affect the quality of the fits, but only influence the final $\chi^2/\text{d.o.f.}$. As mentioned above, we obtained a small $\chi^2/\text{d.o.f.}$ value. For the $\bar{K}^0 \Xi^0$ at $m_\pi=490$ MeV, the scattering length also exhibits considerable uncertainty ($\sim 30\%$). In this case, the LSp fit result shows deviation from lattice QCD, while the iterative LSu fit provides improved agreement within the error range of the lattice QCD. However, it still maintains some distance from the central value of the lattice QCD. This suggests that improving the description of this data point by increasing the order of the chiral expansion is quite challenging. This situation calls for more precise lattice QCD data to resolve this difference.

\subsection{Bayesian statistical analysis}
Bayes's theorem provides a principled framework to update the probability distribution of parameters $\vec{\alpha}$ using data $D$ and prior information $I$. The posterior distribution $\text{pr}(\vec{\alpha}|D,I)$ is given by
\begin{align}
\label{beyesi}
    \text{pr}(\vec{\alpha}|D,I)=\frac{\text{pr}(D|\vec{\alpha},I)\cdot \text{pr}(\vec{\alpha}|I)}{\text{pr}(D|I)},
\end{align}
where the likelihood $\text{pr}(D|\vec{\alpha},I)$ measures how well the parameters explain the data, while $\text{pr}(\vec{\alpha}|I)$ represents prior knowledge about the parameters. The denominator $\text{pr}(D|I)$ ensures proper normalization. In this work, we construct the parameter posterior distribution using lattice QCD data $D$ and prior information $I$. Since the denominator $\text{pr}(D|I)$ is independent of the parameters $\vec{\alpha}$, it can be omitted for parameter estimation and we therefore have
\begin{align}
\label{beyesisimple}
    \text{pr}(\vec{\alpha}|D,I)\propto \text{pr}(D|\vec{\alpha},I)\cdot\text{pr}(\vec{\alpha}|I).
\end{align}
The Bayesian approach provides a principled framework for explicitly incorporating prior information $I$, offering computational transparency.

Here we specify the prior distributions for the LECs in our chiral expansion analysis:
\begin{align}
\label{beyesiconst}
\vec{\alpha}_{\text{NLO}} &= (b_D,b_F,b_0,b_1,b_2,b_3,b_4) \quad [\text{GeV}^{-1}], \nonumber\\
\vec{\alpha}_{\text{NNLO}} &= (\bar{h}_1,\bar{h}_{123}) \quad [\text{GeV}^{-2}], \nonumber\\
\vec{\alpha}_{\text{NNNLO}} &= (\bar{e}_1,\bar{e}_2) \quad [\text{GeV}^{-3}].
\end{align}
Given that the LECs are expected to obey natural size, where $\vec{\alpha}_{\text{NLO}} \sim 1/\Lambda_\chi$, $\vec{\alpha}_{\text{NNLO}} \sim 1/\Lambda_\chi^2$, and $\vec{\alpha}_{\text{NNNLO}} \sim 1/\Lambda_\chi^3$, we consequently adopt constrained prior distributions by directly specifying the range $(-10,10)$ for all LECs. Additionally, for the parameter $\mu\,[\text{GeV}]$ used in our iterative fitting procedure, we establish a prior range of $(0.5,1)$ based on preliminary least-squares fit results.

\begin{table*}[!t]
\centering
\begin{threeparttable}
\caption{\label{tab:consBSfitting}Values of the various fits through the Bayesian statistical analysis. The Mean, StdDeV, Median, and $68\%$CI denote the mean value, standard deviation, median value, and $68\%$ credible interval of the parameters, respectively. The ESS and ESS/s present the effective sample size and the effective sample size per second of compute time, respectively. The $\hat{R}$ is potential scale reduction factor on split chains (at convergence, $\hat{R}=1$). For a more detailed description of these fits, see the main text.}
\renewcommand{\arraystretch}{1.3}
\begin{tabular}{cccccccc}
\toprule \toprule
 Fit BSp & Mean & StdDev & Median & $68\%$CI & ESS & ESS/s & $\hat{R}$ \\
\midrule

$b_D\,(\text{GeV}^{-1})$&$0.12$&$0.27$&$0.16$&[$-0.15$,\,\,$0.39$]&$6415$&$2.24$&$1.002$\\$b_F\,(\text{GeV}^{-1})$&$\hspace{-0.7em}-0.15$&$0.20$&$\hspace{-0.7em}-0.14$&[$-0.34$,\,\,$0.05$]&$8448$&$2.95$&$1.001$\\$b_0\,(\text{GeV}^{-1})$&$1.87$&$0.94$&$2.02$&[$0.92$,\,\,$2.80$]&$5823$&$2.04$&$1.002$\\$b_1\,(\text{GeV}^{-1})$&$1.89$&$1.87$&$2.16$&[$0.01$,\,\,$3.72$]&$5612$&$1.96$&$1.002$\\$b_2\,(\text{GeV}^{-1})$&$\hspace{-0.7em}-0.72$&$0.93$&$\hspace{-0.7em}-0.71$&[$-1.65$,\,\,$0.20$]&$7679$&$2.68$&$1.001$\\$b_3\,(\text{GeV}^{-1})$&$5.56$&$3.06$&$6.05$&[$2.45$,\,\,$8.70$]&$6680$&$2.34$&$1.001$\\$b_4\,(\text{GeV}^{-1})$&$\hspace{-0.7em}-1.27$&$1.88$&$\hspace{-0.7em}-1.48$&[$-3.12$,\,\,$0.58$]&$5223$&$1.83$&$1.002$\\$\bar{h}_1\,(\text{GeV}^{-2})$&$4.93$&$2.30$&$5.17$&[$2.58$,\,\,$7.31$]&$5440$&$1.90$&$1.002$\\$\bar{h}_{123}\,(\text{GeV}^{-2})$&$2.88$&$3.32$&$3.11$&[$-0.45$,\,\,$6.24$]&$5317$&$1.86$&$1.002$\\$\bar{e}_1\,(\text{GeV}^{-3})$&$\hspace{-0.7em}-0.92$&$2.68$&$\hspace{-0.7em}-1.06$&[$-3.64$,\,\,$1.77$]&$6352$&$2.22$&$1.002$\\$\bar{e}_2\,(\text{GeV}^{-3})$&$\hspace{-0.7em}-4.98$&$3.05$&$\hspace{-0.7em}-5.33$&[$-8.14$,\,\,$-1.89$]&$5532$&$1.93$&$1.002$\\

\midrule
\midrule

 Fit BSu & Mean & StdDev & Median & $68\%$CI & ESS & ESS/s & $\hat{R}$ \\
\midrule

$b_D\,(\text{GeV}^{-1})$&$1.16$&$0.63$&$1.16$&[$0.52$,\,\,$1.80$]&$14330$&$6.78$&$1.000$\\$b_F\,(\text{GeV}^{-1})$&$\hspace{-0.7em}-0.47$&$0.65$&$\hspace{-0.7em}-0.46$&[$-1.13$,\,\,$0.18$]&$15340$&$7.26$&$1.000$\\$b_0\,(\text{GeV}^{-1})$&$0.98$&$1.26$&$1.17$&[$-0.40$,\,\,$2.25$]&$15570$&$7.37$&$1.000$\\$b_1\,(\text{GeV}^{-1})$&$\hspace{-0.7em}-0.37$&$3.28$&$\hspace{-0.7em}-0.43$&[$-3.69$,\,\,$2.92$]&$15620$&$7.39$&$1.000$\\$b_2\,(\text{GeV}^{-1})$&$\hspace{-0.7em}-4.18$&$2.44$&$\hspace{-0.7em}-4.14$&[$-6.73$,\,\,$-1.72$]&$14150$&$6.69$&$1.001$\\$b_3\,(\text{GeV}^{-1})$&$3.16$&$4.85$&$4.00$&[$-2.17$,\,\,$8.23$]&$15970$&$7.56$&$1.000$\\$b_4\,(\text{GeV}^{-1})$&$6.59$&$2.58$&$7.13$&[$3.94$,\,\,$9.14$]&$23310$&$11.03$&$1.000$\\$\bar{h}_1\,(\text{GeV}^{-2})$&$6.60$&$2.51$&$7.11$&[$4.02$,\,\,$9.11$]&$19210$&$9.09$&$1.000$\\$\bar{h}_{123}\,(\text{GeV}^{-2})$&$\hspace{-0.7em}-3.21$&$4.04$&$\hspace{-0.7em}-3.61$&[$-7.49$,\,\,$1.08$]&$21470$&$10.16$&$1.000$\\$\bar{e}_1\,(\text{GeV}^{-3})$&$2.45$&$5.15$&$3.29$&[$-3.46$,\,\,$7.90$]&$24830$&$11.75$&$1.000$\\$\bar{e}_2\,(\text{GeV}^{-3})$&$\hspace{-0.7em}-5.56$&$2.98$&$\hspace{-0.7em}-6.05$&[$-8.55$,\,\,$-2.58$]&$22490$&$10.64$&$1.000$\\$\mu\,(\text{GeV})$&$0.90$&$0.07$&$0.91$&[$0.83$,\,\,$0.97$]&$19070$&$9.02$&$1.001$\\

\bottomrule \bottomrule
\end{tabular}
\end{threeparttable}
\end{table*}

\begin{figure}[!t]
\centering
\includegraphics[height=14.30cm,width=13.5cm]{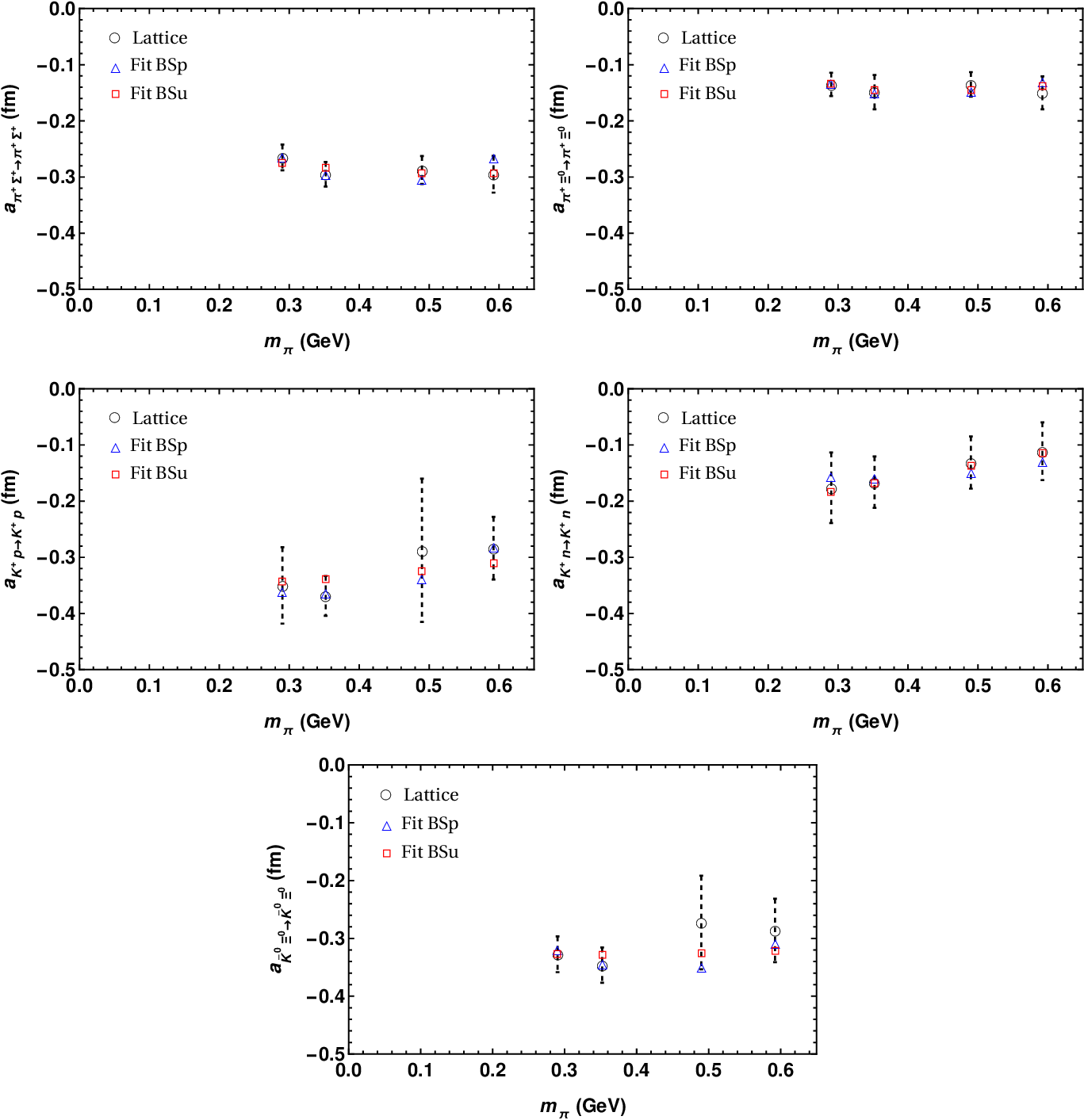}% Here is how to import EPS art
\caption{\label{fig:fitBS}Comparison between the five-channel fitting results obtained using the Bayesian statistical analysis and the lattice QCD data for scattering lengths reported in ref.~\cite{Torok:2009dg}.}
\end{figure}

We condition our LEC posterior on the Lattice QCD data from ref.~\cite{Torok:2009dg}. The corresponding numerical values are selected to be consistent with those from the least squares fit. In this work, we establish the relationship between the lattice QCD value $a_\text{lat}$ and the theoretical value $a_\text{th}$ of the scattering length $a$ through a statistical model:
\begin{align}
\label{beyesimodel}
    a_\text{lat}=a_{\text{th}}+\delta a_{\text{lat}},
\end{align}
where the lattice uncertainty $\delta a_\text{lat}$ is treated as a random variable, for which we directly adopt both the systematic and statistical errors from ref.~\cite{Torok:2009dg} as its quantitative measures. The lattice QCD error for each datum is assumed to follow a normal distribution with variance $(\sigma^2_{\text{sta}}+\sigma^2_{\text{sys}})/3$, i.e.,
\begin{align}
\label{beyesidis}
     \text{pr}(\delta a_\text{lat}|I)=\mathcal{N}(0,(\sigma^2_{\text{sta}}+\sigma^2_{\text{sys}})/3),
\end{align}
where the factor of $1/3$ arises from converting a uniform distribution of errors into a normal distribution. Then the likelihood for a single scattering length reads
\begin{align}
\label{beyesidisreg}
    \text{pr}(a_{\text{lat}}|\vec{\alpha},\sigma,I)=\mathcal{N}(a_{\text{th}}(\vec{\alpha}),(\sigma^2_{\text{sta}}+\sigma^2_{\text{sys}})/3).
\end{align}

We perform Bayesian inference via Markov Chain Monte Carlo (MCMC) sampling using the Stan statistical platform \cite{Stan2024}. Our analysis compares two distinct approaches: Fit BSp incorporates a complete perturbative treatment through Eq.~(\ref{eq:pscatteringlength}), while Fit BSu adopts a non-perturbative framework based on Eq.~(\ref{eq:iscatteringlength}). The resulting LECs from both fitting procedures are presented in Table~\ref{tab:consBSfitting}. For both fits, we run 8 chains with 10000 iterations per chain (5000 warmup and 5000 regular draws). We can observe that all parameters have $\hat{R}$ values very close to one, indicating our model has achieved excellent convergence. The ESS values for all parameters are above 5000, showing we get enough samples. Based on ESS and $\hat{R}$ values, the non-perturbative fitting (Fit BSu) show higher sampling efficiency and better convergence than the perturbative fitting (Fit BSp). This finding is consistent with the smaller $\chi^2$ values observed in non-perturbative fitting compared to perturbative fitting under the least squares method. For these two fits, the mean and median of the parameters are approximately the same, indicating that their distributions are largely symmetric. However, for some parameters (e.g., $b_0$, $b_1$), there are still slight differences, suggesting that their distributions are not perfectly symmetric. Therefore, in the subsequent predictions of physical quantities, we adopt the median (central value) rather than the mean. From the standard deviations and credible intervals, we can see that all these parameters have large uncertainties, which agrees with the least squares results. The main reason is that these parameters are highly correlated with each other. As a result, the scattering length predicted using these parameter distributions will not have significant errors. In principle, we could orthogonalize these parameters, meaning we recombine them to obtain new parameters that would not have such large uncertainties. However, this transformation would not affect the final prediction of the scattering length. Therefore, we do not perform this procedure in the present work.

The corresponding scattering lengths from the two fits are presented in Fig.~\ref{fig:fitBS}. The median values of the posterior scattering lengths obtained from these two fitting methods show excellent agreement with the lattice QCD results in all five channels. For the $K^+p$ channel at $m_\pi \simeq 490$ MeV with substantial lattice uncertainties, our results closely match the lattice central values and agree with the least squares method. In the case of the $\bar{K}^0\Xi^0$ channel at $m_\pi\simeq 490$ MeV which also has significant lattice errors, the Fit BSp shows some deviation while the Fit BSu produces results closer to yet still somewhat different from the central values. These observations agree with the least squares results, indicating that simply changing fitting methods is unlikely to improve the outcomes, while simultaneously verifying the reliability of our fits. Future progress will ultimately depend on achieving higher precision in lattice calculations.

\subsection{Scattering lengths at physical point}
We can now use the determined parameters to predict the scattering lengths at the physical point. For the hadron masses at the physical point, we employ the following values: $m_\pi=139.57\,\text{MeV}$, $m_K=493.68\,\text{MeV}$, $M_N=938.92\,\text{MeV}$, $M_\Sigma=1191.01\,\text{MeV}$, $M_\Xi=1318.26\,\text{MeV}$ \cite{ParticleDataGroup:2024cfk}. For the decay constants at the physical point, we use: $f_\pi=92.21\,\text{MeV}$, $f_K=111.03\,\text{MeV}$ \cite{ParticleDataGroup:2024cfk}. First, let us examine the results obtained using perturbative methods, as shown in Table~\ref{tab:scalthp}. We observe that the least-squares LSp fit fails to achieve complete convergence in all five channels. For the $\pi^+\Sigma^+$ and $\pi^+\Xi^0$ channels, the leading-order contribution dominates while the second- and third-order contributions remain small. However, the fourth-order contribution becomes significantly larger, indicating poor convergence within the SU(3) framework at the fourth order. This suggests that pion-baryon scattering may require an SU(2) framework for proper convergence, as the intermediate kaon states potentially disrupt chiral convergence.

\begin{table}[!t]
\centering
\begin{threeparttable}
\caption{\label{tab:scalthp}Predictions of the scattering lengths
through the perturbative formula, Eq.~(\ref{eq:pscatteringlength}). The scattering lengths are in units of
fm.}
\renewcommand{\arraystretch}{1.3}
\begin{tabular}{cccccc}
\midrule \toprule
Fit LSp & $\mathcal{O}(p)$ & $\mathcal{O}(p^2)$ & $\mathcal{O}(p^3)$ & $\mathcal{O}(p^4)$ & \text{Total}  \\
\midrule
$a_{\pi^{+}\Sigma^{+}}$ &$-0.23$& $-0.02$ &$\hspace{-0.7em}-0.03$ &$0.15$& $-0.13(4)$  \\

$a_{\pi^{+}\Xi^{0}}$ &$-0.12$& $-0.02$ & $\hspace{-0.7em}-0.05$ & $0.15$ & $-0.05(2)$  \\

$a_{K^{+}p}$ &$-0.41$& $-0.14$  & $0.19$  & $\hspace{-0.7em}-0.07$ &$-0.43(8)$\\

$a_{K^{+}n}$ &$-0.21$& $-0.15$   & $0.50$   & $\hspace{-0.7em}-0.34$ & $-0.20(6)$ \\

$a_{\bar{K}^0\Xi^0}$ &$-0.46$& $-0.16$   & $0.25$   & $0.10$ & $-0.26(3)$  \\

\midrule
\midrule 

Fit BSp & $\mathcal{O}(p)$ & $\mathcal{O}(p^2)$ & $\mathcal{O}(p^3)$ & $\mathcal{O}(p^4)$ & \text{Total}  \\
\midrule
$a_{\pi^{+}\Sigma^{+}}$ &$\hspace{-0.7em}-0.23$&$\hspace{-0.7em}-0.04$&$\hspace{-0.7em}-0.03$&$0.16$&$\hspace{-0.7em}-0.14^{+0.02}_{-0.03}$\\

$a_{\pi^{+}\Xi^{0}}$ &$\hspace{-0.7em}-0.12$&$\hspace{-0.7em}-0.04$&$\hspace{-0.7em}-0.05$&$0.15$&$\hspace{-0.7em}-0.06(2)$\\

$a_{K^{+}p}$ &$\hspace{-0.7em}-0.41$&$\hspace{-0.7em}-0.25$&$0.37$&$\hspace{-0.7em}-0.07$&$\hspace{-0.7em}-0.37(7)$\\

$a_{K^{+}n}$ &$\hspace{-0.7em}-0.21$&$\hspace{-0.7em}-0.24$&$0.62$&$\hspace{-0.7em}-0.31$&$\hspace{-0.7em}-0.14(5)$\\

$a_{\bar{K}^0\Xi^0}$ &$\hspace{-0.7em}-0.46$&$\hspace{-0.7em}-0.28$&$0.45$&$0.00$&$\hspace{-0.7em}-0.28(3)$\\
\bottomrule \midrule
\end{tabular}
\end{threeparttable}
\end{table}

Similarly, for the kaon-baryon scattering channels, while the fourth-order contribution is smaller than the third-order, the latter exceeds the second-order contribution. The convergence behavior is also not good in kaon-baryon channels. This implies that complete convergence in these channels might require calculations extended to even higher orders. However, all our final results yield negative values, consistent with physical expectations.

The Bayesian BSp fit demonstrates remarkable agreement with the least-squares results, particularly for $\pi^+\Sigma^+$ and $\pi^+\Xi^0$ where the numerical values coincide precisely. For kaon-baryon scattering channels, some differences emerge, though all cases maintain the same pattern where the third-order exceeds the second-order while the fourth-order becomes smaller again. Particularly in $K^+p$ and $\bar{K}^0\Xi^0$ channels, the fourth-order contributions become negligible. The final results remain consistent with least-squares predictions within error margins. In the Bayesian fits, while the posterior distributions of the LECs exhibit some degree of asymmetry, the posterior distributions of the scattering lengths demonstrate remarkable symmetry. Except for a slight asymmetry in the uncertainties of the $\pi^+\Sigma^+$ channel, all other channels show perfectly symmetric error distributions.

In brief, our perturbative approach for predicting scattering lengths at the physical point exhibits non-convergent behavior across channels, with all final results being physically reasonable. The convergence patterns observed suggest that the SU(3) chiral expansion for all channels may require either higher-order calculations or alternative frameworks like SU(2) for optimal convergence.

Given the suboptimal convergence behavior of pure perturbative theory, we can employ an iterative approach to make non-perturbative predictions for the scattering lengths. Let us now examine the non-perturbatively predicted scattering lengths, with the results presented in Table~\ref{tab:scatteringlength}. We observe that the fitting results from the least squares method (LSu) and the Bayesian approach (BSu) are essentially consistent. The Bayesian fit for the $\pi^+\Sigma^+$ channel is marginally smaller than the least-squares result but still agrees within uncertainties. The errors in Fit BSu are predominantly symmetric, with slight asymmetries only in $\pi^+\Xi^0$, $K^+p$, and $K^+n$. In any case, the non-perturbative predictions for the scattering length at the physical point all yield negative values, which is physically reasonable.

\begin{table}[!t]
\centering
\begin{threeparttable}
\caption{\label{tab:scatteringlength}Predictions of the scattering lengths at the physical point through various methods. The scattering lengths are in units of fm.}
\renewcommand{\arraystretch}{1.3}
\begin{tabular}{ccccccccccccccccccc}
\midrule \toprule
  &$a_{\pi^{+}\Sigma^{+}}$ & $a_{\pi^{+}\Xi^{0}}$ & $a_{K^{+}p}$ & $a_{K^{+}n}$ & $a_{\bar{K}^0\Xi^0}$ & \\
\midrule
Fit LSu &$-0.15(4)$& $-0.05(3)$ &$-0.39(3)$ &$-0.23(5)$& $-0.33(3)$ \\

Fit BSu &$-0.21(2) $&$-0.04^{+0.04}_{-0.03}$&$-0.35^{+0.01}_{-0.02}$&$-0.22^{+0.04}_{-0.03}$&$-0.32(1)$ \\

Fit LSp & $-0.13(4)$& $-0.05(2)$ & $-0.43(8)$ & $-0.20(6)$ & $-0.26(3)$ \\

Fit BSp & $-0.14^{+0.02}_{-0.03}$ & $-0.06(2)$ & $-0.37(7)$ & $-0.14(5)$ & $-0.28(3)$\\

Median & $-0.16(7)$ & $-0.04(4)$ & $-0.41(11)$ & $-0.19(10)$ & $-0.30(7)$ \\

SU(3) NNLO \cite{Torok:2009dg} & $-0.197(08)$ & $-0.096(12)$ & $-0.154(63)$ & $0.128(87)$ & $-0.127(70)$\\

SU(2) NNLO \cite{Torok:2009dg} & $-0.197(08)$ & $-0.098(12)$ & & &\\

Martin \cite{Martin:1980qe} & & & $-0.33$ &  $-0.16$ &  &\\

Meng \cite{Meng:2003gm} & & & $-0.17\pm0.59$ &   &  &\\

Liu (HB) \cite{Liu:2007ct} & $-0.35^{+0.004}_{-0.002}$ & $-0.18$ & $-0.33$ & $-0.16$ & $-0.28^{+0.10}_{-0.05}$ &\\

Mai (HB) \cite{Mai:2009ce} & $-0.24^{+0.01}_{-0.01}$ & $-0.17^{+0.03}_{-0.03}$ & $-0.33^{+0.10}_{-0.10}$ & $-0.16^{+0.14}_{-0.14}$& $-0.33^{+0.11}_{-0.11}$ & \\

Mai (IR) \cite{Mai:2009ce} & $+0.01^{+0.04}_{-0.04}$ & $+0.02^{+0.06}_{-0.07}$ & $-0.33^{+0.32}_{-0.32}$ & $-0.16^{+0.36}_{-0.36}$& $-0.26^{+0.34}_{-0.34}$ & \\

Lu (EOMS) \cite{Lu:2018zof} & &  & $-0.328(5)$ & $-0.170(3)$&  & \\

\bottomrule \midrule
\end{tabular}
\end{threeparttable}
\end{table}

We also include the scattering lengths at the physical point predicted by perturbative methods in Table~\ref{tab:scatteringlength} for comparison. It can be observed that the scattering length values for all five channels are consistent within their respective error ranges. We then extract the median values from these four different fitting approaches and conservatively estimate their uncertainties by taking the maximum deviation between the median value and the extreme values. By comparing with the third-order (NNLO) predictions from ref.~\cite{Torok:2009dg}, we find that for both pion-baryon scattering channels, our results agree with their values within uncertainties, whether in the SU(3) or SU(2) framework. However, for the three kaon-baryon scattering channels, we notice that the third-order predictions appear less reasonable, particularly the $K^+n$ channel which yields a positive value. Our results are consistent, within uncertainties, with the model-dependent extractions of kaon-nucleon scattering lengths from experimental scattering data \cite{Martin:1980qe}. The isospin $I=1$ kaon-nucleon scattering length was computed with results showing agreement with our current prediction in ref.~\cite{Meng:2003gm}. Furthermore, our results for the kaon-baryon scattering lengths show consistency with other HBChPT findings, including both the decuplet-incorporated calculations by Liu et al. \cite{Liu:2007ct} and the truncated-treatment results from Mai et al. \cite{Mai:2009ce}. They are also consistent with the results from the relativistic frameworks of IR \cite{Mai:2009ce} and EOMS \cite{Lu:2018zof}. However, the use of the IR framework yields a positive result for the pion-baryon scattering lengths, which may be attributed to the need for higher-order calculations in the IR framework for pion-baryon scattering. Overall, based on input from lattice data combined with our scattering length formalism, we have obtained reasonable predictions for the physical scattering lengths in all five annihilation-free channels.

Finally, it should be noted that the contributions from decuplet baryons have not been included in our current work. In fact, such contributions could be significant for several reasons. First, the mass difference between decuplet and octet baryons, $\delta= 294$ MeV, is relatively small. Second, the coupling constant between decuplet and octet baryons and pseudoscalar mesons is large. As a result, including decuplet states may partially cancel some intermediate octet contributions. However, incorporating decuplet contributions at the fourth order of chiral expansion would introduce additional LECs. Given the current limitations of available input data, it is difficult to reliably determine these extra LECs. Furthermore, we believe that the LECs used in our current framework are already sufficient to effectively absorb possible contributions from decuplet states. Additionally, our iterative formula, Eq.~(\ref{eq:iscatteringlength}), is not designed to handle coupled-channel effects involving decuplet states. A proper treatment of such effects would require using the original Lippmann–Schwinger equation. Therefore, we have chosen not to include decuplet contributions in this study. We may revisit this aspect in future work when more sufficient input data become available.

\section{Summary}
\label{summary}
In this work, we have comprehensively calculated the threshold $T$ matrices of the meson and baryon processes that have no annihilation diagrams: $\pi^+\Sigma^+$, $\pi^+\Xi^0$, $K^+p$, $K^+n$, and $\bar{K}^0\Xi^0$ to the fourth order using the heavy baryon chiral perturbation theory. By performing least squares and Bayesian fits to the non-physical lattice QCD data, we determined the LECs through both perturbative and non-perturbative iterative methods. This led to a good description of the scattering lengths at most of the
lattice QCD data points. The physical scattering lengths were obtained by extrapolating the corresponding parameters to their physical values. We found that the convergence behavior is not good across all channels in the perturbative method. We obtained the scattering lengths for the five channels by taking the median values from four different fitting approaches: $a_{\pi^+\Sigma^+}=-0.16\pm 0.07\,\text{fm}$, $a_{\pi^+\Xi^0}=-0.04\pm0.04\,\text{fm}$, $a_{K^+p}=-0.41\pm 0.11\,\text{fm}$, $a_{K^+n}=-0.19\pm 0.10\,\text{fm}$, and $a_{\bar{K}^0\Xi^0}=-0.30\pm 0.07\,\text{fm}$, where the uncertainties were conservatively estimated by taking the maximum deviation between the median values and the extreme values of the statistical errors from the four fits. Our calculations for these five channels without annihilation diagrams show clear improvements compared to purely third-order results, and we look forward to future experimental or lattice QCD confirmation of these predictions.

\section*{Acknowledgments}
This work is supported by the National Natural Science Foundation of
China under Grants No. 12075126, No. 12147127, and No. 12565019. B.-L. Huang is also supported by the Inner Mongolia Autonomous Region Natural Science Fund (No. 2024MS01016), the Research Support for High-Level Talent at the Autonomous Region Level (No. 13100-15112042) and the Junma Program High-Level Talent Reseach Start-up Fund (No. 10000-23112101/085).

\clearpage
\appendix\markboth{Appendix}{Appendix}
\renewcommand{\thesection}{\Alph{section}}
\numberwithin{equation}{section}
\section{Parameters of $\mathcal{T}_{\phi B}^{(\text{N3LO})}$ }
\label{parameters}
In this appendix, we present the parameters  $\alpha_i$, $\beta_i$, $\gamma_i$, $\delta_i$ and $\epsilon_i$ ($i=1, 2, 3, 4, 5$) for the $T$-matrices $\mathcal{T}_{\phi B}^{(\text{N3LO})}$ derived from loop-diagram contributions at the fourth order, where the index $(i=1,2,3,4,5)$ corresponds to the ($\pi^+\Sigma^+$, $\pi^+\Xi^0$, $K^+p$, $K^+n$, $\bar{K}^0\Xi^0$) channels, respectively. The relevant parameters for each channel are given by
\begin{align}
\label{a1}
    \alpha_1=&\frac{1}{2304}(-64b_0-512b_D-42 b_3+567 b_4+246 b_1)+\frac{1}{144}(16b_0+32b_D-4 b_3-5 b_4-8 b_1)D^2\nonumber\\
    &+\frac{1}{16}(16b_0+16b_D-4 b_3-3 b_4-4 b_1)F^2,
\end{align}
\begin{align}
\label{a2}
    \beta_1=\frac{1}{384}(32b_0+16b_D-2 b_3-b_1),
\end{align}
\begin{align}
\label{a3}
    \gamma_1=&\frac{1}{36}(4b_0-9b_D)+\frac{1}{144}(104b_0+140b_D-26 b_3-22 b_4-35 b_1)D^2+\frac{1}{8}(4b_F- b_2)DF\nonumber\\
    &+\frac{1}{16}(8b_0+12b_D-2 b_3-2 b_4-3 b_1)F^2,
\end{align}
\begin{align}
\label{a4}
    \delta_1=\frac{1}{4}(-2b_D+ b_4+b_1)m_\pi^3-\frac{1}{2}b_D m_\pi m_K^2,
\end{align}
\begin{align}
\label{a5}
    \alpha_2=&\frac{1}{2304}(-64b_0+192b_D+704b_F-42 b_3-144 b_4-165 b_1-411 b_2)+\frac{1}{288}(8b_0+4b_D\nonumber\\
    &-4b_F-2 b_3-b_1+ b_2)(13D^2-30DF+9F^2),
\end{align}
\begin{align}
\label{a6}
    \beta_2=\frac{1}{768}(64b_0+48b_D+16b_F-4 b_3- b_4-3 b_1- b_2),
\end{align}
\begin{align}
\label{a7}
    \gamma_2=&\frac{1}{72}(8b_0+17b_D+35b_F)+\frac{1}{288}(136b_0+16b_D-128b_F-34 b_3+9 b_4-4 b_1+32 b_2)D^2\nonumber\\
&+\frac{1}{48}(40b_0-32b_F-10 b_3+3 b_4+8 b_2)DF+\frac{1}{32}(40b_0+16b_D-32b_F-10 b_3\nonumber\\
&+ b_4-4 b_1+8 b_2)F^2,
\end{align}
\begin{align}
\label{a8}
    \delta_2=\frac{1}{4}(2b_F- b_2)m_\pi^3+\frac{1}{2}b_F m_\pi m_K^2,
\end{align}
\begin{align}
\label{a9}
    \epsilon_2=\frac{1}{8}(-2b_D-2b_F+ b_4+b_1+ b_2)m_\pi^3-\frac{1}{4}(b_D+b_F)m_\pi m_K^2,
\end{align}
\begin{align}
\label{a10}
    \alpha_3=\frac{1}{13824}(1344b_0+960b_D+192b_F-84 b_3-39 b_1-23 b_2),
\end{align}
\begin{align}
\label{a11}
    \beta_3=&\frac{1}{6912}(864b_0-1872b_D-4464b_F-168 b_3+1701 b_4+705 b_1+41 b_2)+\frac{1}{288}(136b_0\nonumber\\&+204b_D-36b_F-34 b_3-32 b_4-51 b_1+9 b_2)D^2-\frac{1}{48}(40b_0+52b_D-12b_F-10 b_3\nonumber\\&-8 b_4-13 b_1+3 b_2)DF+\frac{1}{32}(40b_0+44b_D-4b_F-10 b_3-8 b_4-11 b_1+ b_2)F^2,
\end{align}
\begin{align}
\label{a12}
    \gamma_3=&-\frac{1}{432}(24b_0+90b_D-282b_F-3 b_3-3 b_1+ b_2)+\frac{1}{288}(104b_0+140b_D+36b_F\nonumber\\&-26 b_3-22 b_4-35b_1-9 b_2)D^2+\frac{1}{48}(40b_0+52b_D+12b_F-10 b_3-8 b_4\nonumber\\&-13 b_1-3 b_2)DF+\frac{1}{32}(8b_0+12b_D+4b_F-2 b_3-2 b_4-3 b_1- b_2)F^2,
\end{align}
\begin{align}
\label{a13}
    \delta_3=\frac{1}{16}(-20b_F+3 b_4+2 b_2)m_K^3+\frac{3}{4}b_F m_K m_\pi^2,
\end{align}
\begin{align}
\label{a14}
    \epsilon_3=\frac{1}{16}(4b_D-8b_F-3 b_4-2 b_1+4 b_2)m_K^3+\frac{1}{4}(b_D-2b_F)m_K m_\pi^2,
\end{align}
\begin{align}
\label{a15}
    \alpha_4=\frac{1}{13824}(1344b_0+768b_D-84 b_3-53 b_1-37 b_2),
\end{align}
\begin{align}
\label{a16}
    \beta_4=&\frac{1}{6912}(864b_0+4032b_D+1440b_F-168 b_3-441 b_4-541 b_1-1205 b_2)+\frac{1}{288}(136b_0\nonumber\\&+52b_D-92b_F-34 b_3-13 b_1+23 b_2)D^2-\frac{1}{48}(40b_0+28b_D-20b_F-10 b_3\nonumber\\&-7 b_1+5 b_2)DF+\frac{1}{32}(40b_0+20b_D-28b_F-10 b_3-5 b_1+7 b_2)F^2,
\end{align}
\begin{align}
\label{a17}
    \gamma_4=&-\frac{1}{432}(24b_0+120b_D-252b_F-3 b_3-2 b_1+2 b_2)+\frac{1}{288}(104b_0+16b_D\nonumber\\&-88b_F-26 b_3+9 b_4-4 b_1+22 b_2)D^2+\frac{1}{48}(40b_0+8b_D-32b_F-10 b_3\nonumber\\&+3 b_4-2 b_1+8 b_2)DF+\frac{1}{32}(8b_0-8b_F-2 b_3+ b_4+2 b_2)F^2,
\end{align}
\begin{align}
\label{a18}
    \delta_4=\frac{1}{16}(10b_D-10b_F- b_1+ b_2)m_K^3+\frac{3}{8}(-b_D+b_F) m_K m_\pi^2,
\end{align}
\begin{align}
\label{a19}
    \epsilon_4=\frac{1}{16}(2b_D-10b_F- b_1+5 b_2)m_K^3+\frac{1}{8}(b_D-5b_F)m_K m_\pi^2,
\end{align}
\begin{align}
\label{a20}
    \alpha_5=\frac{1}{13824}(1344b_0+960b_D-192b_F-84 b_3-39 b_1+23 b_2),
\end{align}
\begin{align}
\label{a21}
    \beta_5=&\frac{1}{6912}(864b_0-1872b_D+4464b_F-168 b_3+1701 b_4+705 b_1-41 b_2)+\frac{1}{288}(136b_0\nonumber\\&+204b_D+36b_F-34 b_3-32 b_4-51 b_1-9 b_2)D^2-\frac{1}{16}(40b_0+36b_D-4b_F-10 b_3\nonumber\\&-4 b_4-9 b_1+ b_2)DF+\frac{1}{32}(40b_0+44b_D+4b_F-10 b_3-8 b_4-11 b_1- b_2)F^2,
\end{align}
\begin{align}
\label{a22}
    \gamma_5=&-\frac{1}{432}(24b_0+90b_D+282b_F-3 b_3-3 b_1- b_2)+\frac{1}{288}(104b_0+140b_D\nonumber\\&-36b_F-26 b_3-22 b_4-35 b_1+9 b_2)D^2+\frac{1}{16}(40b_0+36b_D+4b_F-10 b_3\nonumber\\&-4 b_4-9 b_1- b_2)DF+\frac{1}{32}(8b_0+12b_D-4b_F-2 b_3-2 b_4-3 b_1+ b_2)F^2,
\end{align}
\begin{align}
\label{a23}
    \delta_5=\frac{1}{16}(20b_F+3 b_1-2 b_2)m_K^3-\frac{3}{4}b_F m_K m_\pi^2,
\end{align}
\begin{align}
\label{a24}
    \epsilon_5=\frac{1}{16}(4b_D+8b_F-3 b_4-2 b_1-4 b_2)m_K^3+\frac{1}{4}(b_D+2b_F)m_K m_\pi^2.
\end{align}

\newpage
\bibliographystyle{utphys}
\bibliography{Meson_Baryon_Scattering_Lengths}% Produces the bibliography via BibTeX.
\end{document}